\documentstyle[12pt,epsf]{article}
\newlength{\mathspace}

\begin{document}
\setlength{\oddsidemargin}{0cm}
\setlength{\baselineskip}{7mm}
\setlength{\mathspace}{2.5mm}

\def\del{\partial}
\def\pbar{\bar p}
\def\xbar{\bar x}
\def\unit{{\bf 1}}
\def\picunit#1{{\bf 1}^{#1}}
\def\dualunit#1{{\bf 1}_{*}^{#1}}
\def\singular{\mbox{\bf S}}
\def\ie{{\em i.e., }}
\def\beq{\begin{equation}}
\def\eeq{\end{equation}}
\def\beqa{\begin{eqnarray}}
\def\eeqa{\end{eqnarray}}
\def\vac#1{|#1\rangle}
\def\upvac#1{|#1, \uparrow \rangle}
\def\fock#1{{\cal F}_#1}
\renewcommand{\thesection}{\arabic{section}\setcounter{equation}{0}}
\renewcommand{\theequation}{\arabic{section}.\arabic{equation}}


\begin{titlepage}

    \begin{normalsize}
     \begin{flushright}
                 UT-Komaba/93-7 \\
                 hep-th/9304039 \\
                 April 1993 \\
                 revised December 1993
     \end{flushright}
    \end{normalsize}
    \begin{LARGE}
       \vspace{1cm}
       \begin{center}
         {$c=1$ String as a Topological Model} \\
       \end{center}
    \end{LARGE}

  \vspace{5mm}

\begin{center}
           Hiroshi I{\sc shikawa}
           \footnote{E-mail address:
              ishikawa@hep1.c.u-tokyo.ac.jp}
                \ \  and \ \
           Mitsuhiro K{\sc ato}
           \footnote{E-mail address:
              katom@tkyvax.phys.s.u-tokyo.ac.jp} \\
      \vspace{4mm}
        {\it Institute of Physics, College of Arts and Sciences} \\
        {\it University of Tokyo, Komaba}\\
        {\it Meguro-ku, Tokyo 153, Japan}\\
      \vspace{1cm}

    \begin{large} ABSTRACT \end{large}
        \par
\end{center}
\begin{quote}
 \begin{normalsize}
\ \ \ \
The discrete states in the $c=1$ string are shown to be
the physical states of a certain topological sigma model.
We define a set of new fields directly from $c=1$ variables, in terms of which
the BRST charge and energy-momentum tensor are rewritten as
those of the topological sigma model.
Remarkably, ground ring generator $x$ turns out to be
a coordinate of the sigma model.
All of the discrete states realize a graded ring which contains ground ring
as a subset.
 \end{normalsize}
\end{quote}

\end{titlepage}
\vfil\eject

\setcounter{footnote}{0}
\section{Introduction}
\ \ \ \
The existence of so called unbroken phase in string theory was conjectured
some time ago. Such a phase is naturally expected to be described by an
appropriate topological field theory \cite{Witten2,Witten3}.

Even in the conventional formulation one might suspect there is some
remnant of the topological degrees of freedom. Since there is no transverse
degree of freedom, two dimensional string theory is a good place to look for
such things.
Actually we already know the physical spectrum of the theory \cite{LZ,BMP}.
There are
infinite numbers of discrete states which are fairly different
from ordinary
particle modes. One may naturally bear an idea that the discrete states are
the very remnant of topological degrees of freedom we are seeking for.

One of the encouraging hints follows from the gauge theory in four
dimension. Let us consider, for instance, the Gupta-Bleuler quantization
of free electromagnetic field.  In this formulation, physical state
is defined
by the condition
$$
(\partial_{\mu}A^{\mu})^{(+)} | phys \rangle = 0,
$$
where $^{(+)}$ stands for the positive frequency part of the operator.
This restricts physical polarization to the transverse directions for the
generic case of momentum eigenstates. For the vanishing four momentum case,
however, there is no restriction on the polarization. Thus we have
``discrete states'' for the longitudinal and scalar modes.
The same kind of states also appear at level 1 in the critical string.
Remarkable thing here is that such constant degrees of
freedom are responsible
for the masslessness of all the components of gauge field. Because in this
covariant gauge there still remains a global symmetry $\delta A^{\mu} =
a^{\mu}$, where $a^{\mu}$ is an arbitrary constant vector,
but the vacuum breaks
the symmetry with $\langle  A^{\mu} \rangle = 0$ and corresponding
Nambu-Goldstone modes are nothing but four components of the gauge field
itself \cite{FP,BW}.
This mechanism was applied to non-abelian case to try to explain the
confinement phase as an unbroken phase of a special gauge
symmetry \cite{Hata}.

The above example suggests that the discrete states in the $c=1$
string are
relevant to the phases in string theory in an analogous way.
For example, ghost number\footnote{Throughout this paper
we use the convention
that the $SL(2,C)$ invariant vacuum has vanishing ghost number.}
one discrete states in relative cohomology
are characterized as Fock states which are simultaneously singular
and cosingular vectors with respect to the matter part of the
Virasoro algebra. This implies that we have similar kind of states
also in higher dimensional string including the critical one.
Of course, in higher dimensional case they are buried in the
mass shell, \ie usual particle modes with particular momenta.
But we may consider conversely that such siblings of the discrete states
determine the mass shell that possible particle modes belong to.
Actually Lorentz invariance and the existence of a particular state
on a certain mass shell are enough to show the existence of a particle
on that shell. In two dimensional or $c=1$ case,
only a massless ``tachyon''
is a particle mode so that the role of the discrete states in
unbroken phase may be rather clear if any.
For the $c=1$ string, level of discrete states is not restricted to $1$ as in
the critical
case but extends over all the positive integers. Higher level states
have non-vanishing momenta and have not been given a similar
interpretation as done for the level $1$ or vector field case. We need
a more unified view of the discrete states if we consider them as a
clue to the phases in string theory.
Anyway, it is desired to understand the discrete states from such a
viewpoint, and the present paper is, hopefully, the first step toward
this end.

In this paper we will show that the discrete states in the $c=1$
string are understood as physical states of a certain topological
model. Starting from the $c=1$ string we define new fields in terms of
which we can rewrite the BRST charge of the string theory as that of the
topological sigma model. There, notably enough, the ground ring
generator \cite{Witten} $x$
plays a role of scalar field of the sigma model (or we should call it
a coordinate of a kind of topological string).
This relationship is in the operator level not merely in the level of
amplitudes as is in $c<1$ string \cite{Verlinde}.
And all the discrete states are reconstructed
through the BRST cohomology of the topological model. They naturally
realize a graded ring which contains ground ring as a subset.

It has been argued by several groups that the $c=1$ string can be regarded
as a topological model, {\it e.g.,} topological sigma model
\cite{Horava}, twisted $SL(2,R)/U(1)$ model \cite{MV} and
$SL(2,R)/SL(2,R)$ model \cite{G/G}. However, the previous attempts were
based on comparison of the physical spectrum for topological models
with that for the $c=1$ string. It was not clear why the physical
spectrum of strings was reproduced from topological models. Our
approach is more direct in contrast to these. The $c=1$ string can be
viewed as a `bosonization' of the topological sigma
model\footnote{The observation that string theory can be regarded as
bosonization of a topological model was also obtained in
ref.\cite{Distler} for the case of $c=-2$ matter and the
${\hat c} = 1$ fermionic case.},
and the coincidence of the physical spectrums for
both models is immediate. Although an interpretation of $x$ as a
coordinate of a certain three dimensional cone was given in
ref.\cite{Witten}, topological field theoretic nature was not clear in
that time. We emphasize here that the $x$ is directly shown to be a
coordinate of a topological sigma model. This point is also a difference
from ref.\cite{Horava} in which the author started from a topological
model given a priori and tried to describe only a matter part of the $c=1$
theory.

In this sense, our observation may be a good starting point
toward the direct
understanding of the unbroken phase in terms of the field theory
on the world sheet. Namely, unbroken phase is described by a topological
sigma model and broken phase done by conventional string theory.
Alternatively, several backgrounds of strings emerging in the broken phase are
governed by a topological model which is possibly related to the
unbroken phase.

This paper is organized as follows.
In the subsequent section, from the fields of the $c=1$ string
we define new fields such as gauge ghost
and anti-ghost with dimension 0 and 1 respectively and a pair of
scalar fields which become
ingredients of topological sigma model.
In use of the definition of these fields,
the BRST charge and the energy-momentum tensor of the $c=1$ string
surprisingly
turn out to be those of a topological model.
In a sense, the $c=1$ string is a `bosonization' of a topological sigma
model. The Lagrangian and gauge fixing of corresponding topological sigma
model is discussed in section 3.
In section 4 we will show how to understand the discrete state spectrum
from the topological theoretic point of view. All the discrete states
in the $c=1$ string are reproduced by taking into account both the
picture-changed and dual sectors of the topological sigma model. The
final section is devoted to the discussion.

\vspace*{4mm}
\section{Topological description of the $c=1$ string}
\ \ \ \
We begin with rewriting the $c=1$ string in terms of a topological model.
The $c=1$ string is a model which has a two-dimensional target space
with the linear dilaton background. The energy-momentum tensor of the
model takes the form\footnote{We consider the left moving part only.}
\beqa
T(z) &=& T^{M}(z) + T^{G}(z) \nonumber \\
     &=& -\frac{1}{2} (\del X)^2 -\frac{1}{2}(\del \phi)^2
         +\sqrt{2}\, \del^2 \phi - 2b\, \del c - \del b\, c \, ,
  \label{em}
\eeqa
where $X$ and $\phi$ are string coordinates, free bosons with positive
signature, while $b$ and $c$ are the diffeomorphism ghosts. In the
context of non-critical string, $X$ and $\phi$ represent $c=1$ matter
and the Liouville field respectively.
The BRST operator, which governs the physical spectrum of the model,
reads
\beq
  \label{BRST}
  Q = \oint \! dz\; c \left( T^{M} + \frac{1}{2}\, T^{G} \right)
    = \oint \! dz\; (c\, T^{M} + b c \del c) \, .
\eeq
Physical states of the $c=1$ string are obtained as cohomology with
respect to this BRST operator.

Now, we introduce the following set of fields
\beq
  \label{+fields}
  \begin{minipage}{.35\textwidth}
    \vspace{-\abovedisplayskip}
    \begin{eqnarray*}
      B^+ &=& b \, e^{i X^+ } \, ,\\
      C^+ &=& c \, e^{-i X^+ }\, ,
    \end{eqnarray*}
  \end{minipage}
  \begin{minipage}{.35\textwidth}
    \vspace{-\abovedisplayskip}
    \begin{eqnarray*}
      x &=& ( cb + i\, \del X^- ) e^{i X^+} \, ,\\
      \pbar^+ &=& e^{- i X^+ }\, ,
    \end{eqnarray*}
  \end{minipage}
\eeq
where $X^\pm $ means the light-cone combination of the string coordinates
\beq
  X^\pm = \frac{1}{\sqrt{2}} (X \pm i\, \phi) \, .
\eeq
The dimension of these operators is $0$ for $C^+$ and $x$, and $1$ for
$B^+$ and $\pbar^+$,
respectively. Operator product
expansions (OPE) among them are
\beqa
  B^+ (z)\, C^+ (w) &\sim& \frac{1}{z-w} \;\sim\;
   C^+ (z)\, B^+ (w) \, ,\\
  \pbar^+ (z)\, x(w) &\sim& \frac{1}{z-w} \;\sim\;
   -\, x(z)\,\pbar^+ (w)\, .
\eeqa
The OPE of other combinations are regular. Mutual statistics is therefore
fermionic for $B^+, C^+$  and bosonic for $x, \pbar^+$. Note that $C^+$
and $x$ are physical operators, in particular $x$ is one of the ground
ring generators \cite{Witten},
whereas $B^+ $ and $\pbar^+ $ are not physical.

We use these fields to rewrite the $c=1$ string into a topological model;
the former can be viewed as a `bosonization' of the latter. The point is
that the field contents of $B^+, C^+, x$ and $\pbar^+$ are almost the same
as those of a topological model which we discuss in the next section.
Actually, we can regard this system as a topological sigma model with a
pair of complex scalar $x$, $\xbar$ and gauge ghosts $B$, $C$. Necessary
identification is
\beq
  \label{identification}
  B^+ = B\, ,\; C^+ = C\, ,\; \pbar^+ = - \del \xbar \, .
\eeq
One can consider the $c=1$ string as a realization of the system of
fields $B, C, x$ and $\pbar=-\del\xbar$ through a bosonization
(\ref{+fields}). This observation is supported by the following facts.

First, the $SL(2,C)$ vacuum of the $c=1$ string works as the
vacuum of the topological sigma model. This can be seen by introducing
the mode expansion of the operators:
\beq
  \label{mode}
  \begin{minipage}{.35\textwidth}
    \vspace{-\abovedisplayskip}
    \begin{eqnarray*}
      B^{+}_{n} &=& \oint\! dz \; z^n\, B^+(z) \, ,\\
      C^{+}_{n} &=& \oint\! dz \; z^{n-1}\, C^+(z) \, ,
    \end{eqnarray*}
  \end{minipage}
  \begin{minipage}{.35\textwidth}
    \vspace{-\abovedisplayskip}
    \begin{eqnarray*}
      x_{n} &=& \oint\! dz \; z^{n-1}\, x(z) \, ,\\
      \pbar^{+}_{n} &=& \oint\! dz \; z^n\, \pbar^+(z) \, .
    \end{eqnarray*}
  \end{minipage}
\eeq
These modes act on an operator ${\cal O}(z)$ as, for example,
\beq
  B^{+}_{n} {\cal O}(w) = \oint\! dz\, (z-w)^{n}B(z)\, {\cal O}(w) \, ,
\eeq
where the contour is taken to be encircling the point $w$. Acting these
modes on the $SL(2,C)$ vacuum, or equivalently the unit field
$\unit(z)$, we obtain
\beq
  \begin{minipage}{.35\textwidth}
    \vspace{-\abovedisplayskip}
    \begin{eqnarray*}
      B^{+}_{n}\,\unit &=& 0 \;\,\mbox{ for } n \ge 0 \, ,\\
      C^{+}_{n}\,\unit &=& 0 \;\,\mbox{ for } n \ge 1 \, ,
    \end{eqnarray*}
  \end{minipage}
  \begin{minipage}{.35\textwidth}
    \vspace{-\abovedisplayskip}
    \begin{eqnarray*}
      x_{n}\,\unit &=& 0         \;\,\mbox{ for } n \ge 1 \, ,\\
      \pbar^{+}_{n}\,\unit &=& 0 \;\,\mbox{ for } n \ge 0 \, .
    \end{eqnarray*}
  \end{minipage}
\eeq
Gauge ghosts, which have dimension 1 and 0, complex scalar and its
derivative act properly on the vacuum as they should.

Second, making use of the definition (\ref{+fields}), the
energy-momentum tensor of the $c=1$ string turns into that of the
topological sigma model. It can be shown that
\beqa
  T &=& -\frac{1}{2} (\del X)^2 -\frac{1}{2}(\del \phi)^2
         +\sqrt{2}\, \del^2 \phi - 2b\, \del c - \del b\, c \,\nonumber\\
    &=&  i\del X^+ i\del X^- + i\del^2 X^- - i\del^2 X^+
        - 2b\,\del c - \del b\, c \, \\
    &=& \del x \,\pbar^+ - B^+ \del C^+ \, ,\nonumber
\eeqa
where the composite operators, $\del x \,\pbar^+$ and $B^+ \del C^+$, are
defined as the finite part in operator product expansion
\beq
  (AB)(w) = \oint\! dz \frac{1}{z-w} A(z) B(w) \, .
\eeq
By the identification (\ref{identification}) mentioned
before, one can easily recognize that the last expression is the
energy-momentum tensor of the topological sigma model
\beq
  T_{top} = -\del x \,\del \xbar - B \del C \label{Ttop} \, .
\eeq

Third, and most importantly, the BRST operators of both models are
also translated into each other.
The BRST current of the $c=1$ string
gives rise to that of the topological sigma model
\beq
  c\,T^{M} + bc\del c
  = C^+ \del x
  -\del\!\left( c\,i\del X^+ + \frac{1}{2}\del c \right) \, .
\eeq
Total derivative term in the right hand side vanishes upon integration.
Thereby we come to an astonishing result,
\ie BRST operator (\ref{BRST}) of the $c=1$ string is rewritten as
that of the topological sigma model
\beq
  Q_{top} = \oint\! dz \; C \,\del x \label{Qtop} \, .
\eeq

Thus, we have seen
the correspondence of
the energy-momentum tensor and the BRST operator
in both models, which govern physical contents of the model.
This fact suggests a close relationship between the $c=1$ string
and the topological sigma model.
In a sense, the $c=1$ string embodies the topological sigma
model as a kind of bosonization of fields.
Indeed, as we shall show in the later section, the physical
spectrum of the $c=1$ string, in particular the discrete states, can be
characterized as that of the topological model.

Before going into detail, we should remark the $N=2$ superconformal
structure in the $c=1$ string. As is well-known, the topological
sigma model admits a twisted $N=2$ algebra, the generators of which are
\beq
  \label{topN=2}
  \begin{array}{rcl}
      T &=& T_{top} = -\del x \,\del \xbar - B \del C \, ,\\[\mathspace]
      G^+ &=& C \del x \, ,\\[\mathspace]
      G^- &=& - B\del \xbar \, ,\\[\mathspace]
      J &=& C B  \, .
  \end{array}
\eeq
Note that one of the supercurrents $G^+$ is nothing but the BRST current
and the $N=2$ current $J$ is the ghost number current. Since we can
consider these fields as that of the $c=1$ string
through a bosonization (\ref{+fields}), there is also a twisted $N=2$
algebra in the $c=1$ string. Using the explicit
form (\ref{+fields}) of the fields, the $N=2$ generators in the $c=1$
string read
\beq
  \begin{array}{rcl}
       T &=&  {\displaystyle
         -\frac{1}{2} (\del X)^2 -\frac{1}{2}(\del \phi)^2
         +\sqrt{2}\, \del^2 \phi - 2b\, \del c - \del b\, c \, }
          ,\\[\mathspace]
       G^+ &=& {\displaystyle
         c\, T^M + bc\,\del c + \del\!\left(c\, i\del X^+ +
            \frac{1}{2} \del c \right)  \, ,} \\[\mathspace]
       G^- &=& b \, ,\\[\mathspace]
       J &=& cb - i \del X^+ \, .
  \end{array}
\eeq
This time, the supercurrent $G^+$ is not the BRST current itself but
modified with total derivative terms. Clearly, these
terms vanish upon integration and $Q = \oint G^+$ .

We have shown that the existence of the twisted $N=2$ superconformal
algebra (SCA) in the $c=1$ string follows from the fact that the $c=1$
model can be viewed as a bosonization of the topological sigma model in
which the $N=2$ SCA is manifest. In addition to this, we can show there
also exist a series of twisted $N=2$ SCA, with the BRST current as $G^+$,
in the $c=1$ string. The essential point is that we can modify
$G^+$ (\ref{topN=2}) in the topological sigma model by a total
derivative term $\del(C x)$ to yield a series of twisted $N=2$ SCA's,
the generators of which are
\beq
  \begin{array}{rcl}
      T &=& \del x \,\pbar - B \del C \, ,\\[\mathspace]
      G^+ &=& (1 + \lambda) C \del x + \lambda\, \del C\, x \,
        ,\\[\mathspace]
      G^- &=& B\, \pbar \, ,\\[\mathspace]
      J &=& (1 + \lambda) C B  - \lambda\, x \pbar \, ,
  \end{array} \vspace*{\mathspace}
\eeq
where the parameter $\lambda$ is related to the central charge $c$ of the
untwisted algebra
\beq
  c = 3(1 + 2\lambda) \, .
\eeq
Of course, we have the counterpart of
these $N=2$ SCA in the $c=1$ string and one obtains the following result
\beq
  \begin{array}{rcl}
       T &=&  {\displaystyle
         -\frac{1}{2} (\del X)^2 -\frac{1}{2}(\del \phi)^2
         +\sqrt{2}\, \del^2 \phi - 2b\, \del c - \del b\, c \, }
          ,\\[\mathspace]
       G^+ &=& {\displaystyle
          c\, T^M + bc\,\del c + \del\!\left(c\, i\del X^+ +
          \lambda\, c\, i\del X^- +
          \frac{1}{2}(1 + 2\lambda) \del c \right)  \, ,} \\[\mathspace]
       G^- &=& b \, ,\\[\mathspace]
       J &=& cb - (1 + 2\lambda) i\del X^+ +
       \lambda\, i \del X^- \, .
  \end{array}
\eeq
The previous case corresponds to $\lambda = 0$ and has the
untwisted central charge $c=3$.
Since we modify $G^+$ by a total derivative term, the BRST operator is
again obtained as $\oint G^+$.
This modification, which is necessary in order to form the $N=2$
algebra, of the BRST current is essentially the same one as recently
reported, $\lambda=-1$ corresponds to that in ref.\cite{N=2} while
$\lambda=1$ does ref.\cite{MV}.

So far, we consider the topological model consisting of $B^+ ,C^+ ,x$ and
$\pbar^+$. In the $c=1$ string, however, there exists another realization
of the topological sigma model, the fundamental fields of which take the
following form;
\beq
  \label{-fields}
  \begin{minipage}{.35\textwidth}
    \vspace{-\abovedisplayskip}
    \begin{eqnarray*}
      B^- &=& b \, e^{-i X^- } \, ,\\
      C^- &=& c \, e^{i X^- }  \, ,
    \end{eqnarray*}
  \end{minipage}
  \begin{minipage}{.35\textwidth}
    \vspace{-\abovedisplayskip}
    \begin{eqnarray*}
      y &=& ( cb - i\, \del X^+ ) e^{-i X^-} \, , \\
      \pbar^- &=& e^{ i X^- } \, .
    \end{eqnarray*}
  \end{minipage}
\eeq
Again, $y$ is one of the ground ring generators. It is evident that the
argument for the `$+$'-fields is equally applied to these `$-$'-fields.

\vspace*{4mm}
\section{Topological sigma model}
\ \ \ \
In this section we construct a topological sigma model which reproduces
BRST charge and energy-momentum tensor appeared in the previous section.

The basic fields of our model are complex scalar $x(z,\bar z)$ and
$\bar x(z, \bar z)$, and Lagrangian is taken to be ${\cal L}_0 = 0$.
Apparently this theory has a gauge symmetry $\delta x = \epsilon$,
$\delta\bar x = \bar\epsilon$. Physical contents of topological field
theory are encoded in the gauge fixing of the symmetry. According to the
standard procedure \cite{KU} we introduce a pair of ghost $C$, $\bar C$,
anti-ghost $B$, $\bar B$ and Nakanishi-Lautrup field $\lambda$,
$\bar\lambda$, and define nilpotent BRST transformation:
\begin{equation}
  \begin{array}{lclclcl}
  \delta_B x & = & -\bar C & \qquad & \delta_B\bar x & = & -C \\
  \delta_B C & = & 0       & \qquad & \delta_B\bar C & = & 0 \\
  \delta_B B & = & \lambda & \qquad & \delta_B\bar B & = & \bar\lambda\\
  \delta_B\lambda & = & 0  & \qquad & \delta_B\bar\lambda & = & 0
  \end{array}
\end{equation}
where we omit a Grassmann parameter of the transformation.
Then the gauge fixing term of the Lagrangian is given by the BRST
transform which depends on the gauge choice. We adopt the following
gauge fixed Lagrangian:
\begin{eqnarray}
  {\cal L} &=& {\cal L}_0 +
  \delta_B\! \left( -{1\over 2}B\bar\lambda -{1\over 2}\bar B\lambda
  + B\partial_{\bar z}\bar x + \bar B\partial_zx \right)\nonumber\\
  &=& -\lambda\bar\lambda + \lambda\partial_{\bar z}\bar x
  + \bar\lambda\partial_zx + B\partial_{\bar z}C + \bar B\partial_z\bar C\, .
\end{eqnarray}
After integrating out $\lambda$ and $\bar\lambda$, we get
\begin{equation}
  {\cal L} = \partial_zx\partial_{\bar z}\bar x
  + B\partial_{\bar z}C + \bar B\partial_z\bar C \, .\label{L}
\end{equation}

By utilizing the equations of motion, we can separate left mover and
right mover of the fields
\begin{equation}
\label{top_modes}
\begin{array}{lclclcl}
x &=& x(z) + x(\bar z),
& \qquad & \bar x &=& \bar x(z) + \bar x(\bar z),\\
C &=& C(z), & \qquad & \bar C &=& \bar C(\bar z), \\
B &=& B(z), & \qquad & \bar B &=& \bar B(\bar z).
\end{array}
\end{equation}
Then we obtain the energy-momentum tensor and BRST charge for each sector
\begin{eqnarray}
T(z) &=& -\partial_zx(z)\partial_z\bar x(z)-B(z)\partial_zC(z),\label{T}\\
\bar T(\bar z) &=& -\partial_{\bar z}x(\bar z)\partial_{\bar z}\bar
x(\bar z)-\bar B(\bar z)\partial_{\bar z}\bar C(\bar z),\\
Q &=& \oint dz\, C(z)\partial_zx(z), \label{Q}\\
\bar Q &=& \oint d\bar z\, \bar C(\bar z)\partial_{\bar z}\bar x(\bar z).
\end{eqnarray}
These operators are what appeared in the previous section
in rewriting the $c=1$ string.
So, we can say that this topological sigma model accounts for
the physical contents of the $c=1$ string.
More precisely, it is the $p=0$ sector, where $p$ is the momentum
conjugate to $\xbar$, of this model that corresponds to the $c=1$
string, because there is no $\log z$ term in the mode expansion
(\ref{mode}) of $x$.

The Lagrangian (\ref{L}) was considered in a foresighted
paper \cite{Witten2} as a candidate of topological string. At that time
only a vacuum was recognized as a physical state. But more careful
analysis shows that there is a set of zero modes, being physical,
which gives rich structure to the theory. Also this is an essential point
to relate the theory with the conventional string formulation.
Reconstructing physical states of  the $c=1$ string from the view point
of the topological theory will be given in the next section.

\vspace*{4mm}
\section{Discrete states as physical states of a topological model}
\ \ \ \
In this section, we study the physical states of the $c=1$ string in
the light of the topological sigma model. As is well-known \cite{LZ,BMP},
the $c=1$
physical spectrum consists of a massless propagating degree of freedom,
so-called tachyon, and the discrete states which appear at the special
value of momenta. We restrict our attention to the discrete states and
the discrete tachyon; tachyon states with momentum multiple of
$1/\sqrt{2}$ in the $X$ direction.
This choice corresponds to the self-dual radius of $X$.

In the previous sections, we have shown that the $c=1$ string can be
regarded as a bosonization of the topological sigma model. The
energy-momentum tensor and the BRST operator of the topological sigma
model turn into those of the $c=1$ string. Therefore,
one can naturally expect that the physical spectrum of the $c=1$
string can be understood as that of the topological model. As we shall
see, this is the case.

We first examine the physical spectrum of the topological sigma model
which consists of the following fields
\beq
  \label{modes}
  \begin{minipage}{.35\textwidth}
    \vspace{-\abovedisplayskip}
    \begin{eqnarray*}
      B(z) &=& \sum_{n} B_{n} z^{-n-1} \, ,\\
      C(z) &=& \sum_{n} C_{n} z^{-n} \, ,
    \end{eqnarray*}
  \end{minipage}
  \begin{minipage}{.35\textwidth}
    \vspace{-\abovedisplayskip}
    \begin{eqnarray*}
      x(z) &=& \sum_{n} x_{n} z^{-n} \, ,\\
      \pbar(z) &=& \sum_{n} \pbar_{n} z^{-n-1} \, ,
    \end{eqnarray*}
  \end{minipage}
\eeq
where $n$ takes integer value.
The commutation relations among them are
\beq
  \label{comm}
  \{B_m, C_n\} = \delta_{m+n} \,\, ,\,\,
  [x_m, \pbar_n ] = - \delta_{m+n} \, .
\eeq
These modes act on the vacuum $\vac{0}$ as
\beq
  \label{vacuum}
  \begin{minipage}{.35\textwidth}
    \vspace{-\abovedisplayskip}
    \begin{eqnarray*}
      B_{n}\,\vac{0} &=& 0 \;\,\mbox{ for } n \ge 0 \, ,\\
      C_{n}\,\vac{0} &=& 0 \;\,\mbox{ for } n \ge 1 \, ,
    \end{eqnarray*}
  \end{minipage}
  \begin{minipage}{.35\textwidth}
    \vspace{-\abovedisplayskip}
    \begin{eqnarray*}
      x_{n}\,\vac{0} &=& 0         \;\,\mbox{ for } n \ge 1 \, ,\\
      \pbar_{n}\,\vac{0} &=& 0 \;\,\mbox{ for } n \ge 0 \, .
    \end{eqnarray*}
  \end{minipage}
\eeq
The Fock space $\fock{0}$ of the topological model is constructed on
this vacuum by applying the creation modes.

Physical states in this Fock space is defined as cohomology of the
BRST operator $Q$
\beq
  \label{topBRST}
  Q = \oint\! dz\, C(z) \del x(z) = \sum_{n} n C_{n} x_{-n} \, .
\eeq
{}From eq.(\ref{vacuum}), one can see that the vacuum $\vac{0}$ is
annihilated by this BRST operator, $Q \vac{0} = 0$. Hence, the vacuum
is physical, and
the physical spectrum is built on it by applying physical modes.
So, we need to know
which mode becomes physical, or
how the modes (\ref{modes}) are transformed into each
other under the action of the BRST operator. Using the commutation
relations (\ref{comm}), one obtains the following result
\beq
  \label{doublet}
  \begin{array}{ccc}
      B_n \stackrel{Q}{\longrightarrow} & -n\, x_n & \\[\mathspace]
      \mbox{} & \pbar_n & \hspace*{-2mm}
      \stackrel{Q}{\longrightarrow} \hspace*{2mm} -n\, C_n \, .
  \end{array}
\eeq
Note that almost all the modes form doublets under the BRST
transformation, \ie they are unphysical. For $n=0$, however, this doublet
decouples into two singlets and we have four physical modes,
$B_0, x_0, \pbar_0$ and $C_0 $, but two of them annihilate the vacuum
as one can see from eq.(\ref{vacuum}). Therefore we have two physical
modes, $x_0$ and $C_0$, which create the physical states upon the
vacuum $\vac{0}$. The physical spectrum in the Fock space $\fock{0}$
is thus spanned by two kinds of elements
\beq
    x_{0}^{n} \vac{0}  \mbox{\,\, and \,\,} C_{0} x_{0}^{n} \vac{0}
    \,\,\,\, n = 0,1,2,\cdots \, .
\eeq

Next, we examine how these physical states are expressed in the Fock
space of the $c=1$ string.
The $c=1$ string realizes a bosonization of
the topological sigma model
which is carried out by identifying the
fields $B, C, x$ and $\pbar $ with $B^+, C^+, x$ and $\pbar^+$
(See eq.(\ref{+fields})) in the $c=1$ string.
Through this bosonization,
the physical states obtained above are expressed as fields in the
$c=1$ string, where, as was explained in Section 2,
the unit field $\unit(z)$ works as the vacuum.
By simple calculation, we can show that
\beq
  \label{level1}
  \begin{array}{rcl}
  x_{0}\, \unit &=& ( cb + i\, \del X^- ) e^{i X^+} \, ,\\[\mathspace]
  C^{+}_{0}\, \unit &=& c \, e^{-i X^+ } \, , \;\;
  C^{+}_{0} x_{0} \,\unit = c\, i\del X^- + \del c \, .
  \end{array}
\eeq
One can readily see that these indeed belong to the physical spectrum of
the $c=1$ string; $x_{0}\,\unit$ is the ground ring generator, while
$C^{+}_{0}\,\unit$ and $C^{+}_{0} x_{0}\,\unit$ are a discrete
tachyon and one of the level $1$ discrete operators, respectively.

This correspondence of physical spectrum for the topological sigma
model and the $c=1$ string holds in general.
More precisely, we will show that the $c=1$ expressions, $x_{0}^n \,
\unit$ and $C^{+}_{0} x_{0}^{n} \, \unit$, reproduce a part of the
physical spectrum of the $c=1$ string.
Since the translation from the topological sigma model to the $c=1$
string is simply a bosonization, it is clear that these $c=1$
expressions are annihilated by the BRST operator of the $c=1$ string.
Not obvious thing is whether these operators are BRST exact or not.
If it is impossible to write them as BRST exact ones in the $c=1$ Fock
space, they turn out to be the physical operators of the $c=1$ string.
In fact, non-triviality of them can be shown as follows.

Let us suppose that $x_{0}^{m}\,\unit$ is
written as a BRST-exact operator $x_{0}^{m}\,\unit = Q \,\alpha$.
Acting $\pbar^{+}_{0}$ to this operator
and using the commutation relation (\ref{comm}),
one can strip off $x_0$ one by one to reach the vacuum $\unit$. Since the
BRST operator $Q$ commutes with $\pbar^{+}_{0}$, this means that the
vacuum itself is BRST-exact; a wrong statement. Hence, we can conclude
that $x_{0}^{m}\,\unit$ is a non-trivial physical operator.
Non-triviality of the operators $C^{+}_{0} x_{0}^{m}\,\unit $ can be
shown in the same way. This time, we use $B^{+}_{0}$ as an
`annihilation' operator. If $C^{+}_{0} x_{0}^{m}\,\unit $ is
BRST-exact, $x_{0}^{m}\,\unit $ would be also written as a BRST-exact one
by applying $B^{+}_{0}$ to it, which is inconsistent with the previous
result.

Thus, we have shown that the physical states $x_{0}^{m}\,\vac{0}$ and
$C_{0} x_{0}^{m}\,\vac{0} $ of the topological sigma model give rise
to the physical operators of the $c=1$ string through the bosonization
(\ref{+fields}).
The structure of the physical spectrum realized in the $c=1$ Fock space is
depicted in Fig.1.
However, these are not enough to cover all the
physical spectrum of
the $c=1$ string. Both the operators $x_{0}^{m}\,\unit$ and
$C^{+}_{0} x_{0}^{m}\,\unit $ have momentum such as $\exp(in X^+)$,
while there also exist physical states with momentum
$\exp(im X^+ + in X^-)$.
How can we explain these states from the point of view of the
topological model?

In order to answer this question, let us remember that
a bosonized model provides us with much larger Fock space than the
original one has. We can realize several representations with distinct
vacua within this enlarged Fock space. Since the $c=1$ string is a
bosonization for the topological sigma model, the $c=1$ Fock space also
admits representations other than we already
considered, {\em e.g.,} picture-changed one. As we shall show in the
following, it is this
picture-changed representation that generates the rest of the physical
states in the $c=1$ string. So, we turn to the determination of the
physical spectrum of the topological sigma model in the
picture-changed sector.

The picture-changed Fock space $\fock{\theta}$ is constructed on the
picture-changed vacuum $\vac{\theta}$ with bosonic and fermionic
sea level shifted by $\theta$
\beq
  \label{pic_vacuum}
  \begin{minipage}{.35\textwidth}
    \vspace{-\abovedisplayskip}
    \begin{eqnarray*}
     B_{n}\,\vac{\theta} &=& 0 \;\,\mbox{ for } n \ge \theta\,
     ,\\[\mathspace]
     C_{n}\,\vac{\theta} &=& 0 \;\,\mbox{ for } n \ge 1-\theta \, ,
    \end{eqnarray*}
  \end{minipage}
  \begin{minipage}{.35\textwidth}
    \vspace{-\abovedisplayskip}
    \begin{eqnarray*}
     x_{n}\,\vac{\theta} &=& 0 \;\,\mbox{ for } n \ge 1+\theta\,
     ,\\[\mathspace]
     \pbar_{n}\,\vac{\theta} &=& 0 \;\,\mbox{ for } n \ge  \, -\theta \, .
    \end{eqnarray*}
  \end{minipage}
\eeq
The BRST operator $Q$ (\ref{topBRST}) acts on this vacuum as
\beq
  Q \, \vac{\theta} = - \theta \, C_{-\theta} x_{\theta}\, \vac{\theta}\, .
\eeq
The vacuum $\vac{\theta}$ is not physical unless $\theta = 0$, \ie the
canonical sector which we treated before. However, if we saturate this
vacuum with $C_{-\theta}$, we obtain a physical state
\beq
  Q \, C_{-\theta} \vac{\theta} = - C_{-\theta}\, Q\, \vac{\theta} =
   \theta \, C_{-\theta}^2 x_{\theta}\, \vac{\theta}\, = 0 \, .
\eeq
This state is a picture-changed analogue of the `up' vacuum in the
fermionic $BC$-system and we
denote it as $ \upvac{\theta} = C_{-\theta} \vac{\theta}$.
One can construct the physical spectrum in the picture-changed sector
by applying the physical modes on this `up' vacuum $\upvac{\theta}$.

The structure of the spectrum changes drastically whether $\theta$ is
an integer or not. First, we consider the case of fractional $\theta$.
BRST transformation property of the modes is the same as in the canonical
sector $\theta = 0$ (\ref{doublet}). However,
the suffix $n$ of the modes does not take integer value if
$\theta$ is fractional.
All the modes therefore form BRST doublets to leave no physical
modes, since decoupling of the doublets occurs only when $n=0$.
Hence, we have only one physical state $\upvac{\theta}$ in these
sectors.
Next, we treat the case of integer $\theta$. This time, there exist
physical modes $B_0, C_0, x_0$ and $\pbar_0$, and we can construct
physical spectrum by acting these modes on the `up' vacuum
$\upvac{\theta}$. However, two of them annihilate the `up' vacuum
according to the sign of $\theta$. From eq.(\ref{pic_vacuum}) it follows
that
\beq
  \label{phys_up}
  \begin{array}{rcrcl}
      C_0\,\upvac{\theta} &=& \pbar_0\,\upvac{\theta} &=& 0 \;\,\,\,\,
      \mbox{ for } \theta \ge 0 \, ,\\[\mathspace]
      B_0\,\upvac{\theta} &=& x_0\,\upvac{\theta} &=& 0\;\,\,\,\,
      \mbox{ for } \theta < 0 \, .
   \end{array}
\eeq
Therefore, the physical spectrum in the picture-changed sector for
integer $\theta$ consists of the following states
\beq
  \label{picphys}
  \begin{array}{rcrl}
      x_{0}^{n} \upvac{\theta}  &\mbox{and}&
      B_{0} x_{0}^{n} \upvac{\theta} &
      \,\,\,\, n = 0,1,2,\cdots \, \,\,\,\,
      \mbox{ for } \theta = 0,1,2,\cdots \, \\[\mathspace]
      \pbar_{0}^{n} \upvac{\theta} &\mbox{and}&
      C_{0} \pbar_{0}^{n} \upvac{\theta} &
      \,\,\,\, n = 0,1,2,\cdots \, \,\,\,\,
      \mbox{ for } \theta = -1,-2,\cdots \, .
  \end{array}
\eeq

Let us see how these states are expressed in the $c=1$ Fock space through
the bosonization (\ref{+fields}). The picture-changed vacuum $\vac{\theta}$
is realized as a vertex operator which we denote $\picunit{\theta}$
\beq
  \picunit{\theta} = e^{-i\theta X^-} \, .
\eeq
One can easily verify that this operator has the required property
(\ref{pic_vacuum}) as the picture-changed vacuum.
Picture-changed Fock space $\fock{\theta}$ is constructed on this
vacuum by applying the mode operators $B^+_n, C^+_n, x_n$ and
$\pbar^+_n$. Since these modes have momentum such as $\exp(\pm i X^+)$,
momentum of the resulting operators takes the form
$\exp(-i\theta X^- + i n X^+), n=0,\pm 1,\pm 2,\cdots$.
As was noticed in the last
paragraph, $\picunit{\theta}$ is not physical except $\theta =0$.
Physical ground state, the `up' vacuum $\upvac{\theta} = C_{-\theta}
\vac{\theta}$, gives rise to
\beq
  C^+_{-\theta} \picunit{\theta} =
  \oint\! dz\, z^{-\theta -1} C^{+}(z) \picunit{\theta} =
  c\, e^{-i X^{+} - i \theta X^{-}} \, ,
\eeq
which is nothing but the on-shell tachyon vertex operator and clearly
physical in the $c=1$ Fock space.
We have shown that the physical spectrum for the topological
sigma model consists of only the `up' vacuum if $\theta$ is fractional.
On the other hand, it is well-known that the physical states with
momentum $\exp(-i\theta X^- + i n X^+)$ in the $c=1$ string exist only on
the tachyon shell in the case of fractional $\theta$ (no discrete
states). As was seen above, in the context of the topological
sigma model, this fact is naturally explained as missing of the
physical modes, $B^+_0, C^+_0, x_0$ and $\pbar^+_0$, in the fractional
picture sector.
In the case of integer $\theta$, we have physical modes and the
spectrum is built on the `up' vacuum
$c\, e^{-i X^{+} - i\theta X^{-}}$ by applying
$B^+_0, C^+_0, x_0$ and $\pbar^+_0$.
The physical states (\ref{picphys}) give rise to
\beq
  \label{picphys_op}
  \begin{array}{rcrl}
      x_{0}^{n} C^+_{-\theta} \picunit{\theta}  &\mbox{and}&
      B^+_{0} x_{0}^{n} C^+_{-\theta} \picunit{\theta} &
      \,\,\,\, n = 0,1,2,\cdots \, \,\,\,\,
      \mbox{ for } \theta = 0,1,2,\cdots \, \\[\mathspace]
      (\pbar^+_{0})^{n} C^+_{-\theta} \picunit{\theta} &\mbox{and}&
      C^+_{0} (\pbar^+_{0})^{n} C^+_{-\theta} \picunit{\theta} &
      \,\,\,\, n = 0,1,2,\cdots \, \,\,\,\,
      \mbox{ for } \theta = -1,-2,\cdots \, .
  \end{array}
\eeq
We explain this structure of the $c=1$ spectrum with some concrete
examples.

\

\noindent \underline{\em the case of $\theta$ = 1}

The physical ground state is realized by one of the discrete tachyon
\beq
  C^+_{-1}\,\picunit{\theta=1} =
  c\, e^{-i X^{+} - i X^{-}} = c\, e^{-\sqrt{2} i X} \, .
\eeq
Low level states $B_0 \upvac{\theta =1}$ and $x_0 \upvac{\theta =1}$
turn into
\beq
  \begin{array}{rcl}
  B^+_{0}C^+_{-1}\,\picunit{\theta=1} &=&
  -(cb - i\del X^+) e^{-i X^-} \, ,\\
  x_{0}C^+_{-1}\,\picunit{\theta=1} &=& {\displaystyle
    \left( c\, i \del X^+ i \del X^- + \del c\, i \del X^+
    + c\, i \del^2 X^-
    + bc \del c + \frac{1}{2} \del^2 c \right) e^{-i X^-} \, .}
  \end{array}
\eeq
One can readily see these are the ground ring generator, often denoted
as $y$, and a level 2 discrete operator, respectively.
Besides these two, there are infinite numbers of physical operators
$B^+_{0} x_0^n C^+_{-1}\,\picunit{\theta=1}$ and
$x_0^n C^+_{-1} \,\picunit{\theta=1}$.
Since $B^+$ lowers the $bc$-ghost number by one while $x$ leaves it
unchanged, operators  $B^+_{0} x_0^n C^+_{-1}\,\picunit{\theta=1}$ and
$x_0^n C^+_{-1}\,\picunit{\theta=1}$
have $bc$-ghost number 0 and 1, respectively.
Non-triviality of these
operators in the $c=1$ Fock space can be shown in the same way as in
the canonical sector $\theta=0$.
Hence, we can conclude that the physical states
$B_{0} x_0^n \,\upvac{\theta=1}$ in the topological sigma model give
rise to the physical operators with ghost number 0, \ie ground ring
elements, whereas $x_0^n \,\upvac{\theta=1}$ correspond to the discrete
operators with ghost number 1.

It is easy to extend this analysis to other sectors with positive
$\theta$. The physical ground state is realized by the on-shell tachyon
vertex operator. Physical spectrum on it turns into the discrete
operators in the $c=1$ string; ground ring elements and
operators with ghost number 1.

\

\noindent \underline{\em the case of $\theta = -$1}

Momentum of operators in this sector takes the form
$\exp(i X^- + i n X^+)$. Since these momenta correspond to the tachyon
shell, it is expected that the resulting physical operators are
on-shell tachyons. This can be shown explicitly as follows.

The physical ground state $\upvac{\theta = -1}$ turns into
\beq
  C^+_1\,\picunit{\theta = -1} = c\, e^{- i X^+ + i X^-}
      = c\, e^{\sqrt{2} \phi}
\eeq
which is the cosmological constant operator. Physical
spectrum in this sector is again constructed by applying the physical
modes on it. This time, we can use $C^+_0$ and $\pbar^+_0$ as creation
modes (See eq.(\ref{phys_up})). By simple calculation, one obtains
\beq
  \label{theta=-1}
  \begin{array}{rcl}
  (\pbar^{+}_{0})^{n} C^+_1 \,\picunit{\theta= -1} &=&
    c\, e^{- i (n+1) X^+ + i X^-} \, , \\[\mathspace]
  C^+_0 (\pbar^+_{0})^{n} C^+_1 \,\picunit{\theta= -1} &=&
    \del c \,\, c\, e^{- i (n+2) X^+ + i X^-} \, .
  \end{array}
\eeq
The former belongs to the discrete tachyon while the latter
corresponds to its counterpart in the absolute cohomology.

\

\noindent \underline{\em the case of $\theta = -$2}

The $\theta = -1$ case is, in a sense, exceptional, since its Fock
space is embedded entirely within the tachyon shell. Typical case for
negative $\theta$ starts from $\theta = -2$.

Physical ground state is a discrete tachyon
\beq
  C^+_2\,\picunit{\theta = -2} = c\, e^{- i X^+ + 2 i X^-} \, ,
\eeq
and physical spectrum is constructed in the same way as $\theta = -1$;
$(\pbar^{+}_{0})^{n} C^+_2 \,\picunit{\theta= -2}$ and
$C_0 (\pbar^{+}_{0})^{n} C^+_2 \,\picunit{\theta= -2}$.
BRST non-triviality of these operators can be shown by using $B^+_0$
and $x_0$ as annihilation operators. Therefore, these operators are
physical also in the $c=1$ Fock space and reproduce the $c=1$ discrete
operators with $bc$-ghost number 1 and 2.
Explicit form of low-lying operators reads
\beq
  \label{theta-2}
  \begin{array}{rcl}
  \pbar^+_{0}C^+_{2}\,\picunit{\theta=-2} &=&
  -c\, i \del X^+ e^{-2i X^+ + 2i X^-} \, ,\\
  C^+_{0}C^+_{2}\,\picunit{\theta=-2} &=& {\displaystyle
    \left(\frac{1}{2} \del^2\! c\, c - \del c \, c\, i\del X^+
    \right) e^{-2i X^+ + 2i X^-} \, ,}
  \end{array}
\eeq
which are indeed the discrete operators with momentum
$e^{- 2i X^+ + 2 i X^-} = e^{2 \sqrt{2} \phi}$.

We can apply the same argument to other sectors with negative
$\theta$; physical states
$(\pbar_{0})^{n} \upvac{\theta}$ and
$C_0 (\pbar_{0})^{n} \upvac{\theta}$ in the topological sigma model
turn into the $c=1$ discrete operators with $bc$-ghost number 1 and 2,
respectively.

\

We have seen that the physical states of the topological sigma model,
together with its picture-changed sector,
turn into the $c=1$ discrete states.
It is summarized in Fig.2(a) how these sectors are embedded in the $c=1$ Fock
space.
However, a half of the discrete
states is still missing in this description of the $c=1$ string. At
vanishing momentum, for example, we obtained two physical operators, the
unit field $\unit$ and a level 1 discrete operator $C^+_0 x_0 \unit$,
while there exist four physical states in the $c=1$ string; one state
with ghost number 0, two with 1 and one with 2.
This holds also for other momenta. At each momentum, there exist, at
most, two physical states (\ref{picphys}) of the topological sigma
model, while we have four discrete states at each momentum (except the
tachyon shell, on which we have two physical states). In order to
describe the physical spectrum of the $c=1$ string completely (\ie
including the elements of the absolute cohomology) in terms of the
topological model, we have to characterize another half of the spectrum
also as that of the topological model.

This problem is resolved by taking into account dual representations of
the topological sigma model. Let us suppose a Fock space of the
bosonic ghost system. It is well-known that one can not find the dual
vacuum, \ie a state having non-vanishing norm with the vacuum, in this
Fock space itself. We need another Fock space to evaluate, for
example, expectation value of operators.

It is the same situation that we encounter in our analysis of the
topological sigma model.
Within the Fock spaces $\fock{\theta}$ which are constructed on the
vacua $\vac{\theta}$, it is impossible to find the dual vacua
$\vac{\theta^*}$ such that $\langle \theta^* \vac{\theta} \neq 0 $.
However, we are treating this model in the bosonized form; the $c=1$
string. The $c=1$ Fock space is large enough to admit the dual
representation as well as the picture-changed one. The dual vacuum
$\dualunit{\theta}$ conjugate to
$\picunit{\theta} = e^{-i\theta X^-}$ is realized as
\beq
  \dualunit{\theta} = \frac{1}{2} c\,\del c\, \del^2\! c \,
  e^{-2i X^+ + i(\theta + 2) X^-}
\eeq
in the $c=1$ Fock space.
One can construct the dual spectrum on this vacuum, which is
characterized as conjugate to the physical spectrum in the
original sector.
So, a sector dual to the $\theta$-picture, which is located at
$\exp(inX^+ - i\theta X^-)$, has momentum
$\exp(inX^+ + i(\theta + 2) X^-)$.

The conjugate operator to the
physical ground state
$C^+_{-\theta} \picunit{\theta} = c\, e^{-iX^+ - i\theta X^-}$
takes the following form
\beq
  B^+_{-\theta} \dualunit{\theta} =
  c \, \del c \, e^{-i X^+ + i(\theta + 2) X^-} \, ,
\eeq
which is the absolute cohomology counterpart of the on-shell tachyon.
For fractional $\theta$, we have only this operator in the spectrum.
For integer $\theta$, on the other hand, there are infinite series of
the physical operators. Therefore, we have corresponding
conjugate operators in the case of integer $\theta$. From
eq.(\ref{picphys_op}), one can write down them as follows
\beq
  \label{dualstates}
  \begin{array}{rcrl}
      (\pbar^+_0)^n B^+_{-\theta}\dualunit{\theta}  &\mbox{and}&
      C^+_{0} (\pbar^+_0)^n B^+_{-\theta}\dualunit{\theta} &
      \,\,\,\, n = 0,1,2,\cdots \, \,\,\,\,
      \mbox{ for } \theta = 0,1,2,\cdots \, \\[\mathspace]
      x_{0}^{n} B^+_{-\theta}\dualunit{\theta} &\mbox{and}&
      B^+_0 x_{0}^{n} B^+_{-\theta}\dualunit{\theta} &
      \,\,\,\, n = 0,1,2,\cdots \, \,\,\,\,
      \mbox{ for } \theta = -1,-2,\cdots \, .
  \end{array}
\eeq
These are just what we are seeking for.
The physical spectrum (\ref{picphys_op}) together with its dual
(\ref{dualstates}) has the same multiplicity at each momentum as
the discrete states (See Fig.2(b)).
Since they are
physical and mutually independent, we can conclude that they reproduce
all the discrete states of the $c=1$ string in the absolute
cohomology. We confirm this fact by examining the explicit form of the
dual operators (\ref{dualstates}) for several $\theta$.

\

\noindent \underline{\em the case of dual to $\theta = -$2}

First, we take the
case of $\theta = -2$. This sector has the same momentum
$\exp(inX^+)$ as the canonical sector constructed on the unit field
$\picunit{\theta = 0}$.
The dual ground state
$B^+_2\,\dualunit{\theta=-2} = c \, \del c \, e^{-i X^+}$
is the partner of the
discrete tachyon $C^+_0\, \picunit{\theta=0} = c \, e^{-i X^+}$
in the absolute cohomology.
Low-lying operators read
\beq
  \begin{array}{rcl}
  B^+_0 B^+_2\, \dualunit{\theta=-2} &=&
    - c\, i\del X^+ + \del c \, , \\
  x_0 B^+_2\, \dualunit{\theta=-2} &=& {\displaystyle
    c \, \del c (i\del X^+ + i\del X^-) + \frac{1}{2}c\, \del^2\! c \, .}
  \end{array}
\eeq
These are the level 1 discrete operators with the ghost number 1 and
2, respectively. Note that these operators are independent of those we
obtained in the canonical sector (\ref{level1}).
Hence, together with the canonical sector, all the level 1 operators,
the unit field $\unit$, two discrete operators with ghost number 1 and
one discrete operator with ghost number 2, at the vanishing momentum
are reproduced as the physical states of the topological sigma model.

\

\noindent \underline{\em the case of dual to $\theta = -$1}

We next examine the case of
$\theta=-1$. By simple calculation, the explicit form of the dual
operators (\ref{dualstates}) for $\theta=-1$ can be written as
\beq
  \begin{array}{rcl}
  x_{0}^{n} B^+_1 \dualunit{\theta=-1} &=&
    n!\,\, c\,\del c\, e^{i(n-1)X^+ + i X^-} \, , \\[\mathspace]
  B^+_{0} x_{0}^{n} B^+_1 \dualunit{\theta=-1} &=&
    - n!\,\, c\, e^{in X^+ + i X^-} \, ,
  \end{array}
\eeq
which are the discrete tachyons missing in the $\theta=-1$ sector
(\ref{theta=-1}). We obtain all the discrete tachyons on the tachyon
shell $\exp(inX^+ + iX^-)$ considering both sectors; $\theta=-1$ and
its dual.

\

\noindent \underline{\em the case of dual to $\theta $=0}

As the final example, we take the case of $\theta = 0$,
which has the same momentum as the $\theta=-2$
sector. One obtains
\beq
  \begin{array}{rcl}
  \pbar^+_0 B^+_0 \dualunit{\theta=0} &=&
   \del c\,c\, i\del X^+\, e^{-2iX^+ + 2iX^-}   \, , \\
  C^+_0 B^+_0 \dualunit{\theta=0} &=& {\displaystyle
   -\frac{1}{2} \del^2\! c\, \del c\, c\, e^{-2iX^+ + 2iX^-}\, .}
  \end{array}
\eeq
These are again discrete operators with ghost number 2 and 3,
respectively, and reproduce all the level 1 operators at momentum
$e^{-2iX^+ + 2iX^-}=e^{2\sqrt{2} \phi}$
together with the physical operators (\ref{theta-2}) in the $\theta=-2$
sector.

\

We have shown that, through a bosonization, the physical
spectrum of the topological sigma model turns into that of the $c=1$
string by taking into account the picture-changed sectors and their
dual. All the discrete states and the discrete tachyons in the $c=1$
spectrum has been reproduced as the physical states of the topological
model. Our result is summarized in Figs.2 and 3. Fig.2 shows the location of
each sector in the $c=1$ Fock space, while Fig.3 illustrates how
picture-changed sectors and their dual are embedded in the $c=1$ Fock space
to reproduce the spectrum of the discrete states.

Before closing this section, we point out that our description of the
$c=1$ discrete states (\ref{picphys_op}), (\ref{dualstates}) can be
fit into more concise form by utilizing the `$-$'-fields
(\ref{-fields}). Similarly to the `+'-fields,  zero modes
$B^-_0, C^-_0, y_0$ and $\pbar^-_0$ are physical in the $c=1$ Fock
space. However, they do not exist within the Fock space of the
topological sigma model, since we adopt the `+'-fields (\ref{+fields})
in realizing the fields of the topological model. It may be possible that these
physical modes play the role of the `picture-changing'
operators analogous to those in the fermionic string theories
\cite{FMS}.

In fact, we can show that $y_0$ maps the physical operators in the
$\theta$ sector to those in the $\theta + 1$ sector. By straightforward
calculation, one obtains
\beq
  y_0 C^+_{-\theta} \picunit{\theta} =
    (\theta + 1) C^+_{-(\theta + 1)} \picunit{\theta + 1} \, ,
\eeq
which means that the physical ground states
$C^+_{-\theta} \picunit{\theta} = c\, e^{-iX^+ - i\theta X^-}$ are
mapped to that in the adjacent sector by $y_0$.
{}From this fact, it follows that the physical operators
$x_0^n C^+_{-\theta} \picunit{\theta}$ and
$B^+_0 x_0^n C^+_{-\theta} \picunit{\theta}$ in the $\theta$ picture
are also mapped to those in the $\theta + 1$ picture, since $y_0$
commutes with $B^+_0$ and $x_0$ up to BRST-exact term
\beqa
  \{ y_0, B^+_0 \} &=& 0 \, , \\
  \{ y_0, x_0 \} &=&
  \{ Q, \oint\! dz\,\frac{1}{z^2}(-b\,e^{i(X^+ -X^-)}) \} \, .
\eeqa
Similarly, $C^-_0$ relates the ordinary sector with the dual sector.
The physical ground state $C^+_{-\theta} \picunit{\theta}$ is mapped to
the dual ground state $B^+_{\theta + 1} \dualunit{-\theta -1}$
\beq
  C^-_0 C^+_{-\theta} \picunit{\theta} =
    \del c\, c\, e^{-iX^+ - i(\theta - 1)X^-} =
    - B^+_{\theta + 1} \dualunit{-\theta -1} \, ,
\eeq
and the physical operators
$x_0^n C^+_{-\theta} \picunit{\theta}$ and
$B^+_0 x_0^n C^+_{-\theta} \picunit{\theta}$ in the $\theta$ picture
turn into the dual operators
$x_0^n B^+_{\theta + 1} \dualunit{-\theta -1}$ and
$B^+_0 x_0^n B^+_{\theta + 1} \dualunit{-\theta -1}$.

Therefore, together with the result for $y_0$, all the physical operators
with positive Liouville momentum are obtained starting from those in the
canonical sector, $x_0^n \unit$ and $C^+_0 x_0^n \unit$, by successive
application of the picture-changing operators $y_0$ and $C^-_0$.
The expressions (\ref{picphys_op}) and
(\ref{dualstates}) for the discrete operators can be rewritten (modulo
BRST exact terms) into the following form
\beqa
  \mbox{ghost number 0} \;\;\;\; & &
  x^m_0 y^n_0 \,\unit \, , \nonumber \\
  \mbox{ghost number 1} \;\;\;\; & &
  C^{+}_{0}x^{m}_0 y^n_0 \,\unit \, ,\;\;
  C^{-}_{0}x^m_0 y^{n}_0 \,\unit \, , \\
  \mbox{ghost number 2} \;\;\;\; & &
  C^{+}_{0} C^{-}_{0} x^m_0 y^n_0\,\unit \, ,\nonumber
\eeqa
for the operators with positive Liouville momentum.
Another half of the spectrum is obtained by taking the dual of this.
So, we have expressions
\beqa
  \mbox{ghost number 3} \;\;\;\; & &
  (\pbar^+_0)^m (\pbar^-_0)^n \,\dualunit{} \, ,\nonumber \\
  \mbox{ghost number 2} \;\;\;\; & &
  B^{+}_{0}(\pbar^+_0)^{m} (\pbar^-_0)^n \,\dualunit{} \, ,\;\;
  B^{-}_{0}(\pbar^+_0)^m (\pbar^-_0)^{n} \,\dualunit{} \, , \\
  \mbox{ghost number 1} \;\;\;\; & &
  B^{+}_{0} B^{-}_{0}(\pbar^+_0)^m (\pbar^-_0)^n\,\dualunit{}\, .\nonumber
\eeqa
for those with negative Liouville momentum.

\vspace*{4mm}
\section{Discussions}
\ \ \ \
We have shown that the BRST operators and the energy-momentum
tensor of the $c=1$
string are at the same time those of a topological sigma model in which
the ground ring generator $x$ is one of the basic fields.
In a sense, the $c=1$ string can be viewed as a bosonization of the
topological sigma model. As a result,
all the discrete states in the $c=1$ string have been reproduced as
physical states of the topological model by taking into account the
picture-changed sectors of the latter together with
their dual. Corresponding physical operators form a graded ring which
contains the ground ring of ghost number zero sector as a subset.

An important problem is to clarify a nature of the topological
sigma model which accounts for the discrete states in the $c=1$ string.

As is well-known, the $c=1$ string has a rich structure which is realized
by the discrete states with several ghost numbers. The discrete
operators with ghost number $0$ form the ground ring \cite{Witten}
generated by two operators $x$ and $y$, while those with ghost number
$1$ are related to the $w_{\infty}$-currents \cite{KP}. The latter acts on
the ground ring as an area-preserving diffeomorphism of the $xy$-plane
\cite{Witten}.
It seems that
the operators $x$ and $y$ are on the equal footing in the $c=1$ string.
On the other hand,
our topological sigma model utilizes only $x$ as a target space
coordinate. The operator $y$ does not appear in the topological model
itself but is in one of the picture-changed sectors and works as a
picture-changing operator as is pointed out in Section~4.
$x$ and $y$ thus play distinct roles from the point of view of the
topological sigma model.
The status of $y$ should be figured out in
order to understand the meaning of the topological model.

As is mentioned above, the $w_{\infty}$-currents generate an
area-preserving diffeomorphism of the $xy$-plane, which is a symmetry
of the $c=1$ string. If we consider all the elements of the absolute
cohomology we get the full diffeomorphism instead of the
area-preserving one. Then a natural question arises; what in the
topological theory originates such a diffeomorphism?
In order to answer to this question, remember the gauge fixing of the
topological model adopted in Section~3. Before fixing the gauge, the
model possesses large symmetry $\delta x = \epsilon$ where $\epsilon$ is
an arbitrary function. Our gauge fixing kills all the degrees of freedom
but zero mode. So we have arbitrary transformation of $x_0$ as a residual
symmetry. Apparently such transformation could depend on $x_0$ itself.
Thus the model possesses diffeomorphism symmetry of $x_0$:
\beq
\delta x_0 = \epsilon ( x_0 ).
\eeq
Since the operator $x_0$ maps a physical state to a physical one, the
diffeomorphism of $x$ causes mixing of the physical states. The
operator $y_0$ similarly maps a physical state to that in the adjacent
picture. Therefore we can formally say that the diffeomorphism of $y$
induces mixing of the physical states among different pictures.
However, geometrical meaning of this symmetry in the
topological model is rather obscure in contrast with the case of $x$.

Lastly, we comment on the relation of the
topological sigma model to the vacuum of string theories. At the
beginning of this paper, we argued that the `discrete state' in the
critical string can be viewed as carrying information about the
background geometry of strings.
Since our topological model organizes all the discrete states in the
$c=1$ string, it is expected that the topological sigma model is also
related to the background of strings.
In fact, appearance of discrete states is not restricted to the
$c=1$ case but common feature in string theories with two-dimensional
target space, {\it e.g.,} black hole \cite{BH} and $N=1$ fermionic case
\cite{N=1}. It may be possible that discrete states in these cases are also
governed by a topological model. If we can identify these topological
models for several backgrounds with that for the $c=1$ string, our
topological model is, in a sense, universal, and could be
regarded as parametrizing the background of two-dimensional string
theories. The (second-quantized) wave function of the
topological sigma model may play an important role in such a
description of strings.

\newpage

\section*{Figure captions}

\vspace*{4mm}
\noindent
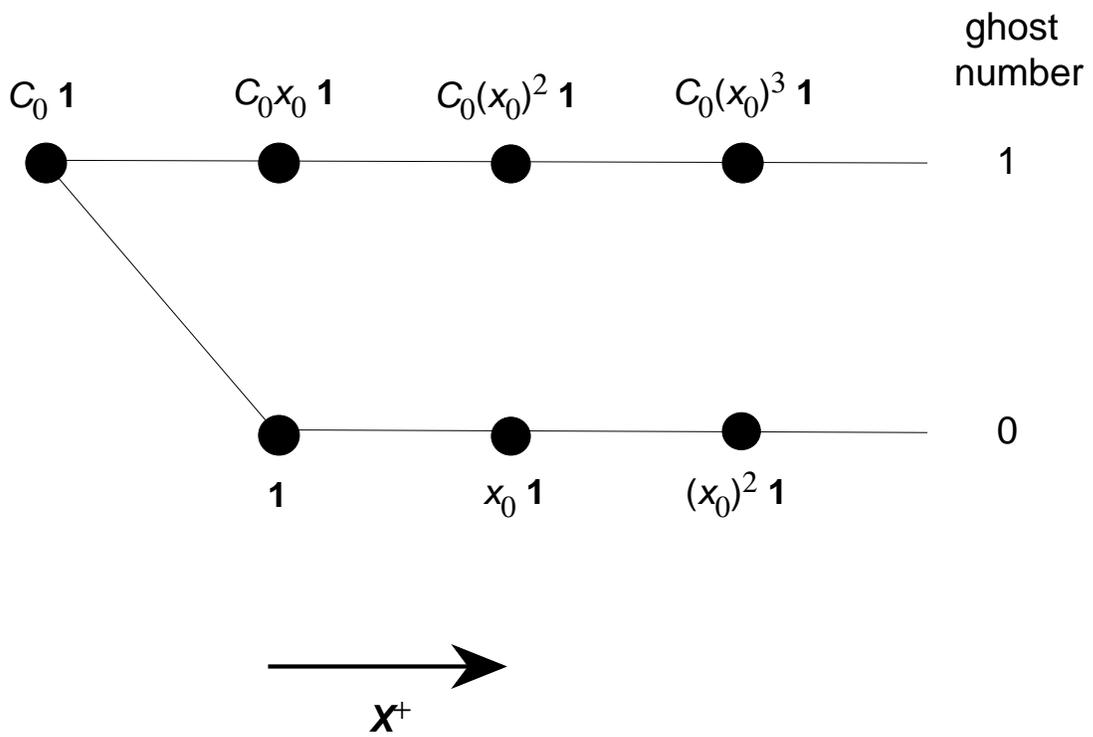
{\bf Fig.1}$\;$
Physical spectrum of the topological sigma model realized in the $c=1$
Fock space.
The horizontal axis represents momentum conjugate to $X^+$
and the vertical axis does $bc$-ghost number.

\

\noindent
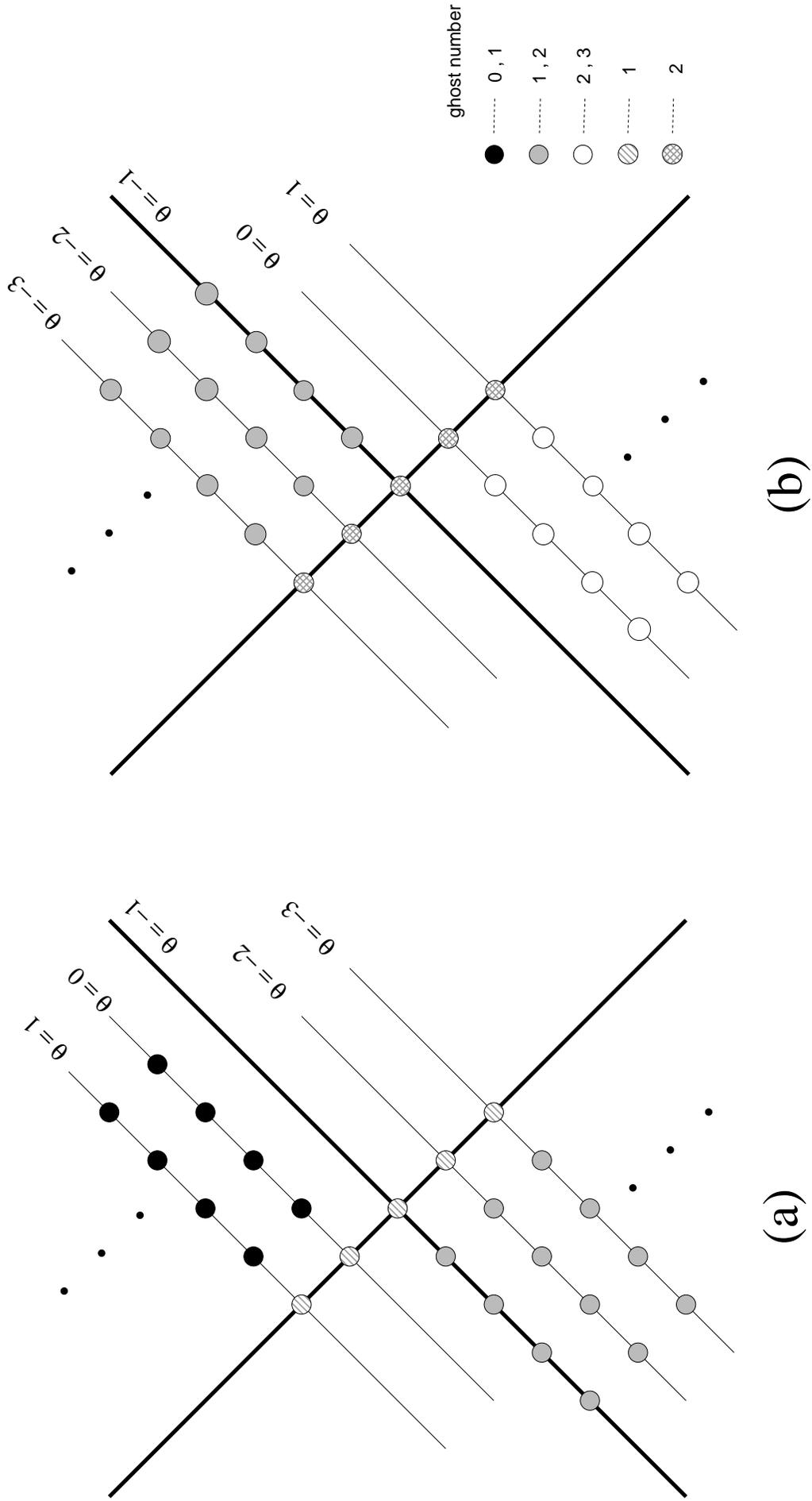
{\bf Fig.2}$\;$
Various sectors of the topological sigma model
realized in the $c=1$ Fock space:(a) picture-changed sectors, (b)dual
sectors.
The solid lines represent the tachyon shell.
Each dot corresponds to the location of the physical states of the
topological model and reproduces all the discrete states in the $c=1$
string.

\

\noindent
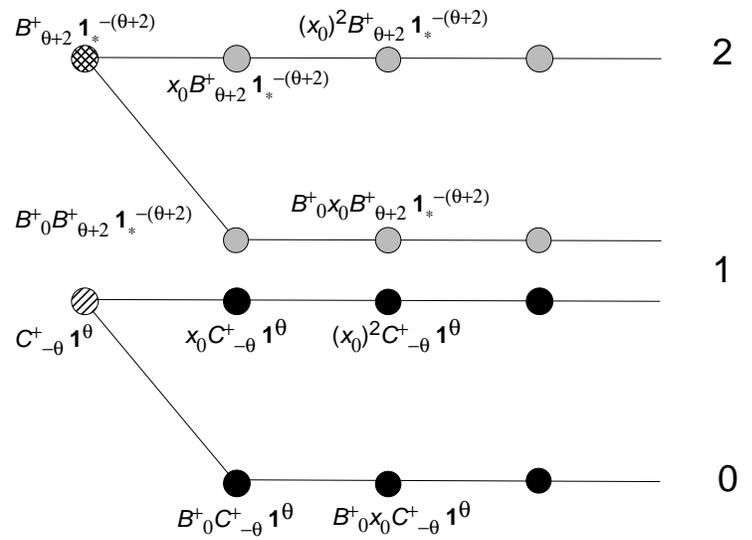
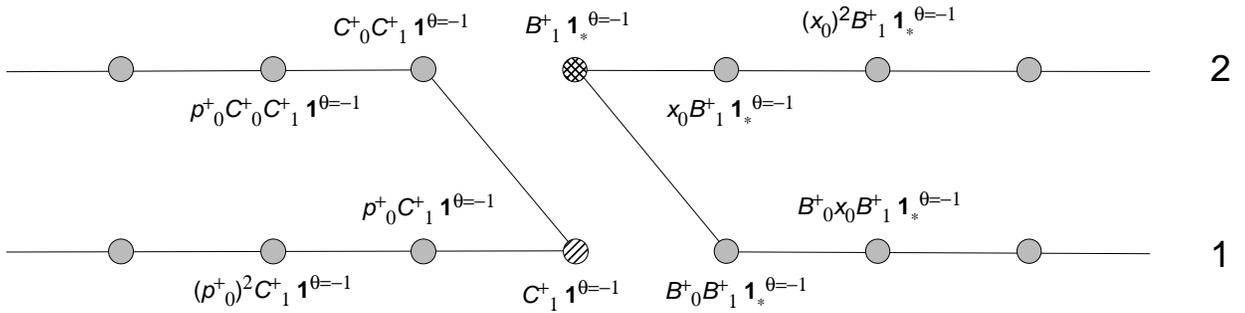
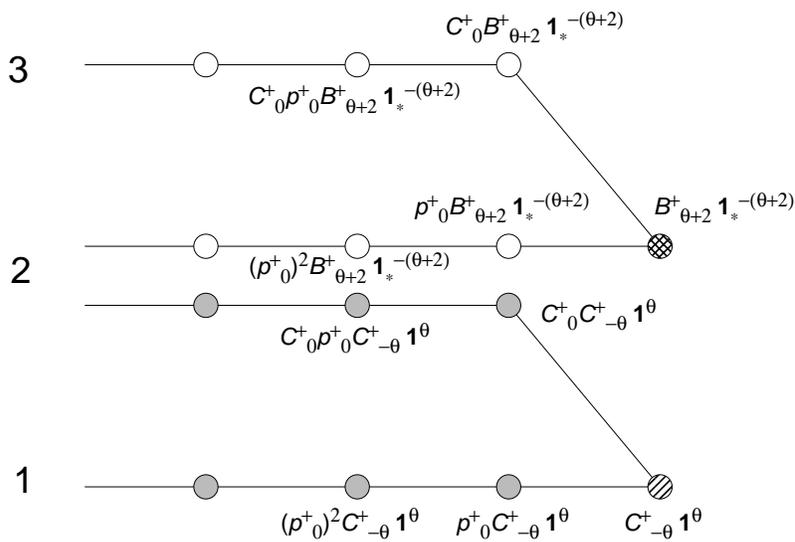
{\bf Fig.3}$\;$
The structure of the physical spectrum of the $c=1$ string in terms of
the topological sigma model. The vertical axis shows the
$bc$-ghost number.
The horizontal direction stands for the momentum conjugate to $X^+$
and corresponds to the solid lines in Fig.2.
Each pattern similar to Fig.1 represents
a spectrum built on one of the vacua. Exactly the same multiplicity of
the discrete states in the $c=1$ string is reproduced.

\newpage
\setlength{\topmargin}{-2.5cm}
\pagestyle{empty}

\vspace*{8cm}
\epsfbox{top_figure1.ai}
\newpage

\vspace*{-1cm}
\epsfbox{top_figure2.ai}
\newpage

\vspace*{-2cm}
\epsfbox{top_figure3.ai}

\end{document}